\documentclass[pra,twocolumn,superscriptaddress,longbibliography]{revtex4-1}   
\usepackage{amsbsy,amssymb,amsmath,bm,bbold} 
\usepackage{graphicx,color,epsfig,rotate} 
\usepackage{fancyhdr} 
\usepackage{epstopdf}
\usepackage{float}
\usepackage[colorlinks=true,linkcolor=blue,citecolor=blue,urlcolor=blue]{hyperref}

\usepackage{tikz}

\usepackage[pagewise]{lineno}

\def\bbbc{{\mathchoice {\setbox0=\hbox{$\displaystyle\rm C$}\hbox{\hbox 
to0pt{\kern0.4\wd0\vrule height0.9\ht0\hss}\box0}} 
{\setbox0=\hbox{$\textstyle\rm C$}\hbox{\hbox 
to0pt{\kern0.4\wd0\vrule height0.9\ht0\hss}\box0}} 
{\setbox0=\hbox{$\scriptstyle\rm C$}\hbox{\hbox 
to0pt{\kern0.4\wd0\vrule height0.9\ht0\hss}\box0}} 
{\setbox0=\hbox{$\scriptscriptstyle\rm C$}\hbox{\hbox 
to0pt{\kern0.4\wd0\vrule height0.9\ht0\hss}\box0}}}}

\DeclareMathAlphabet\mathbfcal{OMS}{cmsy}{b}{n}

\AtBeginDocument{%
	\newwrite\bibnotes
	\def\bibnotesext{Notes.bib}
	\immediate\openout\bibnotes=\jobname\bibnotesext
	\immediate\write\bibnotes{@CONTROL{REVTEX41Control}}
	\immediate\write\bibnotes{@CONTROL{%
			apsrev41Control,author="08",editor="1",pages="1",title="0",year="1"}}
	\if@filesw
	\immediate\write\@auxout{\string\citation{apsrev41Control}}%
	\fi
}%

\begin{document} 
\title{Dynamically Encircled Higher-order Exceptional Points in an Optical Fiber}
\author{Arpan Roy}
\affiliation{Institute of Radio Physics and Electronics, University of Calcutta, Kolkata-700009, India}
\author{Arnab Laha}
\affiliation{Institute of Spintronics and Quantum Information, Faculty of Physics and Astronomy, Adam Mickiewicz University, 61-614 Pozna\'n, Poland}
\author{Abhijit Biswas}
\affiliation{Institute of Radio Physics and Electronics, University of Calcutta, Kolkata-700009, India}
\author{Adam Miranowicz}
\affiliation{Institute of Spintronics and Quantum Information, Faculty of Physics and Astronomy, Adam Mickiewicz University, 61-614 Pozna\'n, Poland}
\author{Bishnu P. Pal}
\affiliation{Department of Physics, \'Ecole Centrale School of Engineering, Mahindra University, Hyderabad-500043, India}
\author{Somnath Ghosh}
\email{somiit@rediffmail.com}
\affiliation{Department of Physics, \'Ecole Centrale School of Engineering, Mahindra University, Hyderabad-500043, India}

\begin{abstract}	
The unique properties of exceptional point (EP) singularities, arising from non-Hermitian physics, have unlocked new possibilities for manipulating light-matter interactions. A tailored gain-loss variation, while encircling higher-order EPs dynamically, can significantly enhance the control of the topological flow of light in multi-level photonic systems. In particular, the integration of dynamically encircled higher-order EPs within fiber geometries holds remarkable promise for advancing specialty optical fiber applications, though a research gap remains in exploring and realizing such configurations. Here, we report a triple-core specialty optical fiber engineered with customized loss and gain to explore the topological characteristics of a third-order exceptional point (EP3), formed by two interconnected second-order exceptional points (EP2s). We elucidate chiral and nonchiral light transmission through the fiber, grounded in second- and third-order branch point behaviors and associated adiabatic and nonadiabatic modal characteristics, while considering various dynamical parametric loops to encircle the embedded EPs. We investigate the persistence of EP-induced light dynamics specifically in the parametric regions immediately adjacent to, though not encircling, the embedded EPs, potentially leading to improved device performance. Our findings offer significant implications for the design and implementation of novel light management technologies in all-fiber photonics and communications.
\end{abstract}  
\maketitle %

\section{Introduction}

Over recent decades, photonic systems have proven to be exceptional platforms for exploring non-Hermitian quantum mechanics (NHQM), primarily due to their intrinsic openness in the sense of ubiquitous loss and gain \cite{Wang23review,Bergholtz2021,Ganainy2019}. In particular, one of the intriguing phenomena in NHQM is the emergence of exceptional points (EPs), a special type of spectral singularities that appears within the parameter space of open systems. At an EP, coupled eigenvalues and their corresponding eigenvectors coalesce simultaneously, creating a defect in the topology of the eigenspace of the underlying Hamiltonian \cite{Kato,Heiss12JPA}. Extensive theoretical and experimental research on implementing EPs across various photonic systems has demonstrated their effectiveness as a powerful tool	for manipulating and detecting the energy states of light \cite{Parto21review,Ozdemir19,Miri19,Ganainy18}. The unique properties of EPs, along with their realization in photonic structures, enable a broad range of advanced quantum-inspired applications, such as topological state-switching \cite{Kullig18njp,Arkhipov2023,Arkhipov2024,Laha21EP4}, asymmetric energy transfer \cite{Xu16,Laha18,Zhang18insitu}, lasing \cite{Peng16pans} and antilasing \cite{Wang21CPA}, slow-light optimization \cite{Goldzak18}, exceptional refrigeration\cite{Lai2024cooling} enhanced nonreciprocity \cite{Choi17isolator,Laha20,Lai2024}, and extremely precise enhanced sensing \cite{Wiersig16,Wiersig20,Hodaei17sensor,Chen17sensor}. Remarkable exploration of EPs has also been observed in quantum optics, paving the way for applications in advanced quantum state engineering \cite{Minganti2019,Minganti2020}, quantum state tomography \cite{Naghiloo2019}, and quantum heat engines \cite{Bu2023}.

Unprecedented system dynamics are observed based on the topological properties of an EP, where chirality plays a key role \cite{Dembowski01}. Quasistatically varying coupling control parameters along a closed loop around an EP results in adiabatic permutations among the corresponding coupled eigenvalues \cite{Dembowski04}. However, the system fails to meet such adiabatic expectations when considering a dynamic effect (time-dependence or analogous length-dependence in photonic systems) in the parametric encirclement process. In this case, regardless of the initial eigenstates, the system ultimately ends up in different particular dominating eigenstates depending on the chirality, in the sense of direction of the EP-encirclement process \cite{Gilary13na,Milburn15na}. Such an intriguing topological property, based on a dynamically encircled EP, enables asymmetric mode conversion in guided-wave geometries, where, regardless of input, light is converted into two distinct dominating modes, while propagating in opposite directions. In the context of EP2s, this phenomenon has been theoretically explored within waveguide \cite{Laha18,Zhang18insitu} and fiber geometries \cite{Roy22} and experimentally validated in a microwave waveguide system \cite{Doppler16}. Furthermore, recent reports have questioned whether it is essential to encircle an EP2 within a parametric loop to achieve asymmetric light dynamics \cite{Hassan2017,Nasari2022}. The findings suggest that similar asymmetric behavior, influenced by both adiabatic and nonadiabatic effects, can also arise when the loop passes near the EP2 without fully enclosing it.

However, investigating complex light behaviors near higher-order EPs is often challenging due to the requirement of an intricate spatial complexity of the underlying photonic system, compounded with increased parametrization \cite{Heiss08JPA,Mandal2021,Sayyad2022}. Nonetheless, recent studies on dynamically encircled higher-order EPs in various waveguide based geometries highlight their great potential for controlling light behavior in multi-level photonic systems \cite{Zhang19EP3,Dey20,Gandhi20,Paul2021}. In this context, optical fiber geometries operating with higher-order EPs hold promise for transformative advances in light guidance schemes, though a significant research gap remains, with only a few reports focusing primarily on EP2s \cite{Roy22,Bergman21}. The main challenge is achieving higher order EPs with minimal parameters, within a single fiber segment, moving beyond the coupled waveguide geometries. To host higher-order EPs, an interesting approach drawing a topological analogy between an EP of order $n$ (EP$n$) and $(n-1)$ interconnected EP2s for an $n$-level system, has been predicted \cite{Muller08,Ryu12} and recently implemented \cite{Laha21EP4,Paul2021}. We apply this approach to optimize a specially designed three-core fiber geometry, aiming to minimize the number of required parameters and constraints.

In this paper, we investigate the topological properties arising from dynamically encircled EP3 formed by two interconnected EP2s within a specially designed three-core optical fiber segment supporting three quasi-guided modes. Non-Hermiticity is attained through a tailored gain-loss profile that is simply modulated across a 2D parameter space, avoiding any need for complex parametrization. We examine the topological behaviors associated with both second- and third-order branch points by exploring various parametric loops in the 2D gain-loss plane, specifically focusing on mode-flipping dynamics. Our primary focus is on the dynamics of light, driven by the asymmetric transfer of modes, while considering the dynamical parametric variations along different loops relative to the locations of the EPs. We particularly emphasize the chiral aspects of the underlying dynamics. Additionally, we highlight a particular case of a dynamical encirclement scheme confined within the interaction regime of embedded higher-order EP, without encircling any of the connected EP2s. The proposed fiber-based dynamical higher-order EP encirclement scheme holds promise for realizing higher-order mode converters with precise mode selectivity, facilitating multi-modal operation in all-fiber networks for advanced communication technologies.

\section{Results and Discussions}

\subsection{Designing the fiber structure}

\begin{figure}[t!]
	\centering
	\includegraphics[width=\linewidth]{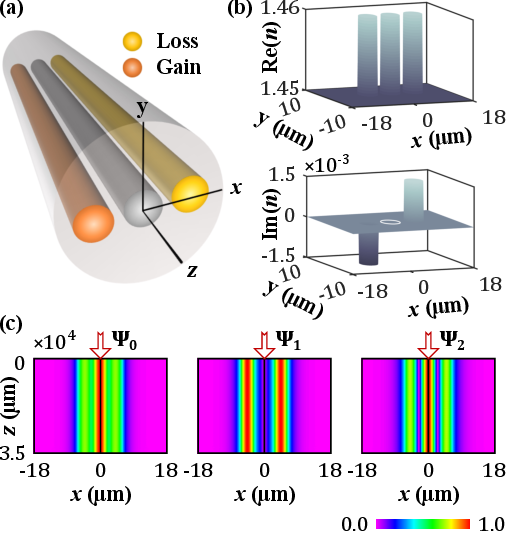}
	\caption{\textbf{(a)} Schematic structural geometry of the proposed fiber segment with three equally sized core ($xy\rightarrow$ transverse plane; $z\rightarrow$ propagation axis). \textbf{(b)} Chosen refractive index profile with the distributions of (upper panel) Re($n$) and (lower panel) Im($n$) for a specific $\gamma=1.2\times10^{-3}$ and $\tau=1$. \textbf{(c)} Beam dynamics of the three supported modes under passive operating conditions (i.e., when $\gamma=0$).}
	\label{fig_1}
\end{figure}

We design a specialty step-index optical fiber segment consisting of three equally-sized cores surrounded by a cladding. Figure \ref{fig_1}(a) shows a schematic illustration of the designed fiber, where the $xy$-plane represents the transverse cross-section, and the z-axis defines the direction of propagation. The refractive indices for the cores ($n_{\text{co}}$) and cladding ($n_{\text{cl}}$), with $n_{\text{co}} > n_{\text{cl}}$, are chosen as $n_{\text{co}} = 1.46$ and $n_{\text{cl}} = 1.45$ to facilitate easier fabrication using silica-based materials. The operating wavelength ($\lambda$) is also fixed at 1.55 $\mu m$ to ensure compatibility with current communication technologies. Given the chosen $n$-values and $\lambda$, we optimize the other geometrical parameters, i.e., $d_{\text{co}}=5\,\mu m$ (core-diameter) and $d=6.7\,\mu m$ (center-to-center separation between the cores), to enable the overall fiber geometry to support three quasi-guided modes: the fundamental mode ($\Psi_0$), the first higher-order mode ($\Psi_1$), and the second higher-order mode ($\Psi_2$). Notably, each individual core can still function as a single-mode fiber under these operating conditions. Along the $z$-axis, the length of the fiber segment is set to $L=35\,mm$.

The designed fiber segment becomes non-Hermitian upon introducing a customized gain-loss profile, where spatially distributed gain is applied to the leftmost core and loss to the rightmost core, while the middle core and cladding remain passive (without any gain-loss). This gain-loss profile is parameterized by two independent tunable (only along $z$-axis) parameters: the gain-loss coefficient ($\gamma$) and the loss-to-gain ratio ($\tau$). Thus, the complex refractive indices of the two outer cores (denoted as $n_\text{L}$ for the left core and $n_\text{L}$ for the right core) can be expressed as
\begin{equation}
n_\text{L}=n_{\text{co}}-i\gamma/\tau\qquad\text{and}\qquad n_\text{R}=n_{\text{co}}+i\gamma, 
\label{eq_sys}
\end{equation}
while the rest of the fiber remains passive throughout the operation. The overall complex refractive index profile, $n(x,y)$, is illustrated in Fig. \ref{fig_1}(b), where the upper and lower panels show the real and imaginary parts of $n(x,y)$, respectively, for a specific case with $\gamma=1.2\times10^{-3}$ and $\tau=1$ (referred to a balanced gain-loss condition). In Fig. \ref{fig_1}(c), we present the beam dynamics of the three supported modes under passive operating conditions (i.e., when $\gamma=0$). However, the introduction gain-loss induces mutual coupling between the quasiguided modes. Such a coupling phenomenon can theoretically be understood by analyzing a three-level perturbed Hamiltonian with an appropriate choice of perturbation elements \cite{Bhattacherjee19_1}.  

In this study, we design the fiber using RSoft\textsuperscript{\textregistered} simulation software. To explore the characteristics of EPs, we analyze the modal properties, where the effective mode indices ($n_{\text{eff}}$) of three quasiguided modes represent the eigenvalues of the underlying Hamiltonian. We apply the finite element method (FEM) to examine transverse modal characteristics, while the beam propagation method (BPM) is used to investigate the modal propagation dynamics.

\subsection{Encountering the multiple EPs and their topological properties}

\begin{figure}[t]
	\centering
	\includegraphics[width=\linewidth]{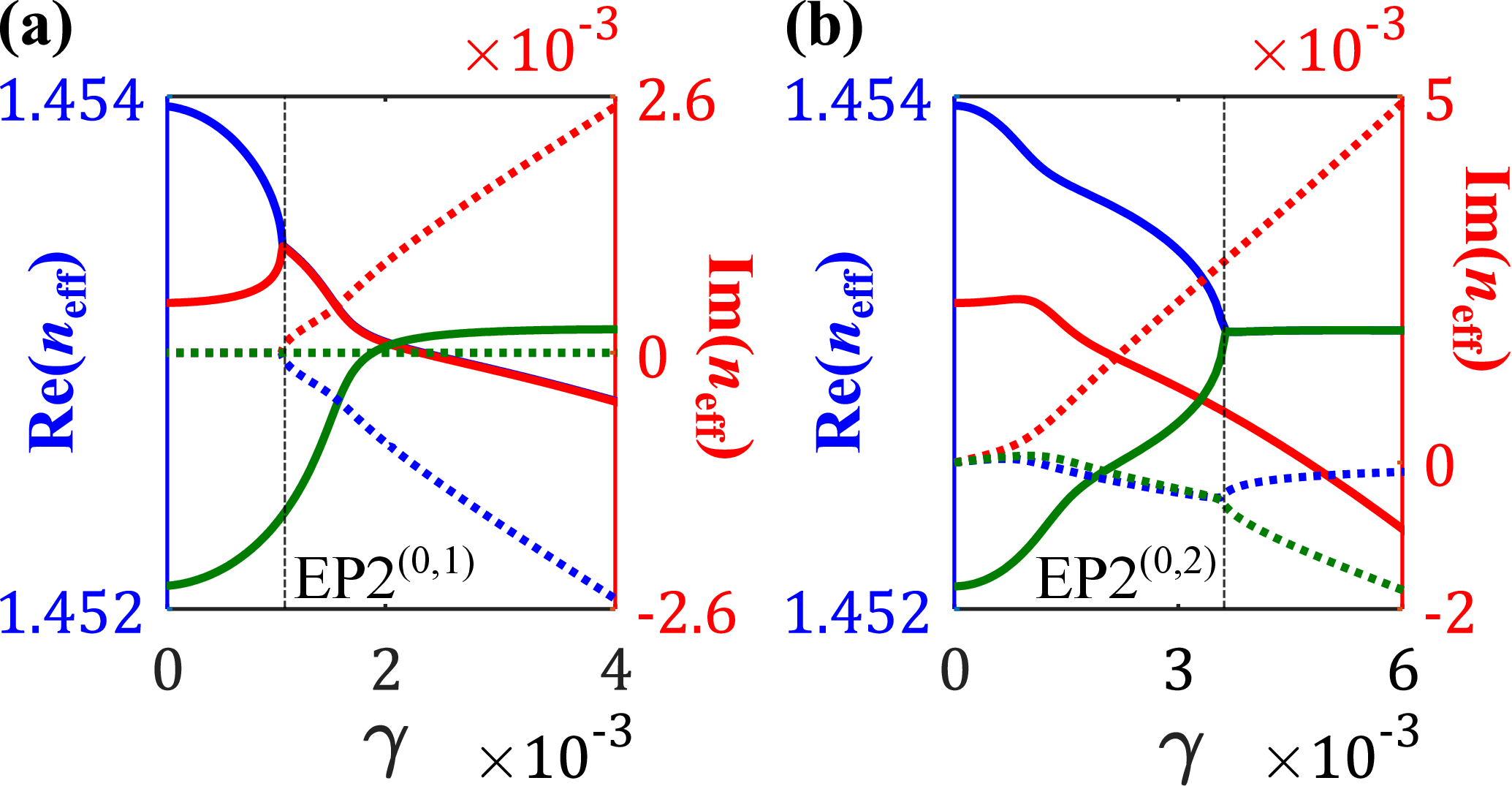}
	\caption{Trajectories of the complex $n_{\text{eff}}$-values associated with $\Psi_0$, $\Psi_1$, and $\Psi_2$ (represented by blue, red, and green lines, respectively) as $\gamma$ increases for two different $\tau$-values. Solid lines indicate the corresponding real part, Re($n_{\text{eff}}$), plotted along the left vertical axis, while dotted lines represent the imaginary part, Im($n_{\text{eff}}$), plotted along the right vertical axis. \textbf{(a)} For $\tau=1$: a simultaneous coalescence of Re($n_{\text{eff}}$) and bifurcation of Im($n_{\text{eff}}$) associated with $\Psi_0$ and $\Psi_1$ occurs at $\gamma=1.1\times10^{-3}$, indicating the emergence of EP2$^{(0,1)}$ in the $(\gamma,\tau)$-plane. \textbf{(b)} For $\tau=2.008$: a simultaneous coalescence of Re($n_{\text{eff}}$) and bifurcation of Im($n_{\text{eff}}$) associated with $\Psi_0$ and $\Psi_2$ occurs at $\gamma=3.6\times10^{-3}$, marking the emergence of EP2$^{(0,2)}$ in the $(\gamma,\tau)$-plane.}
	\label{fig_2}
\end{figure}

In the designed fiber structure, we study the interactions among $\Psi_j\,(j=0,1,2)$ by tracking the trajectories of the associated $n_{\text{eff}}$-values, while varying the gain-loss profile based on the chosen coupling control parameters $\gamma$ and $\tau$. We investigate the avoided-crossing-type interactions among the three modes by varying $\gamma$ within the range $[0,6\times10^{-3}]$ for different $\tau$-values. Through this analysis, we identify two specific cases where the three modes approach two interconnected EP2s, as illustrated in Figs. \ref{fig_2}(a) and \ref{fig_2}(b). The $n_{\text{eff}}$-values associated with $\Psi_j\,(j=0,2,3)$ are color-coded in blue, red, and green, with solid lines representing Re($n_{\text{eff}}$) and dotted lines representing Im($n_{\text{eff}}$) (corresponding to the left and right vertical axes, respectively).

Figure \ref{fig_2}(a) shows the trajectories of $n_{\text{eff}}$ values as $\gamma$ increases for a fixed $\tau=1$. Around $\gamma=1.1\times10^{-3}$, we observe a simultaneous coalescence in Re($n_{\text{eff}}$) and a bifurcation in Im($n_{\text{eff}}$) for the modes $\Psi_0$ and $\Psi_1$. Such a specific type of eigenvalue interactions refers to the presence of an EP2. Thus, we identify an EP2 between $\Psi_0$ and $\Psi_1$ [labeled as EP2$^{(0,1)}$] at $(1.1\times10^{-3},1)$ in the $(\gamma,\tau)$-plane. In this case, the $n_{\text{eff}}$-value associated with $\Psi_1$ moves away from the strong interaction region induced by EP2$^{(0,1)}$. Similarly, for a chosen $\tau=2.008$, we observe another coalescence in Re($n_{\text{eff}}$) and bifurcation in Im($n_{\text{eff}}$) between modes $\Psi_0$ and $\Psi_2$ near $\gamma=3.6\times10^{-3}$, as shown in Fig. \ref{fig_2}(b). This signifies the emergence of another EP2 between $\Psi_0$ and $\Psi_2$ [labeled as EP2$^{(0,2)}$] at $(3.6\times10^{-3},2.008)$ in the $(\gamma,\tau)$-plane. Here, the $n_{\text{eff}}$ value associated with $\Psi_1$ moves away from the strong interaction region induced by EP2$^{(0,2)}$.

Therefore, we observe two distinct scenarios in our three-level system, where modes (hence, $n_{\text{eff}}$-values) from two different pairs coalesce at two EP2s, while the third mode remains unaffected, acting as an observer. These two embedded EP2s are interconnected through $\Psi_0$ under the chosen setup. Such an interaction scheme indicates the emergence of an EP3 \cite{Muller08} within the chosen parametric regime, with its topological properties accessible through the two interconnected EP2s.

Now, we consider different parametric loops in the $(\gamma,\tau)$-plane to study the topological effects induced by the parametrically encircled EPs. By allowing $\gamma$ and $\tau$ to vary according to
\begin{equation}
\gamma(\theta)=\gamma_{0}\sin(\theta/2)\quad\text{and}\quad\tau (\theta)=\tau_{0}+r\sin(\theta),
\label{eq_enc}
\end{equation}
we analyze different encirclement schemes in the $(\gamma,\tau)$-plane. Here, a sufficiently slow angular variation of $\theta$ over the interval $[0,2\pi]$ enables a stroboscopic encirclement scheme. The parameters $\gamma_0$, $\tau_0$, and $r$ (preferably, $r<1$) determine the number of EP2s that can be enclosed within the loop. For a given EP to be properly enclosed within a loop, $\gamma_0$ must exceed the $\gamma$-value associated with that EP.
\begin{figure}[t!]
	\centering
	\includegraphics[width=\linewidth]{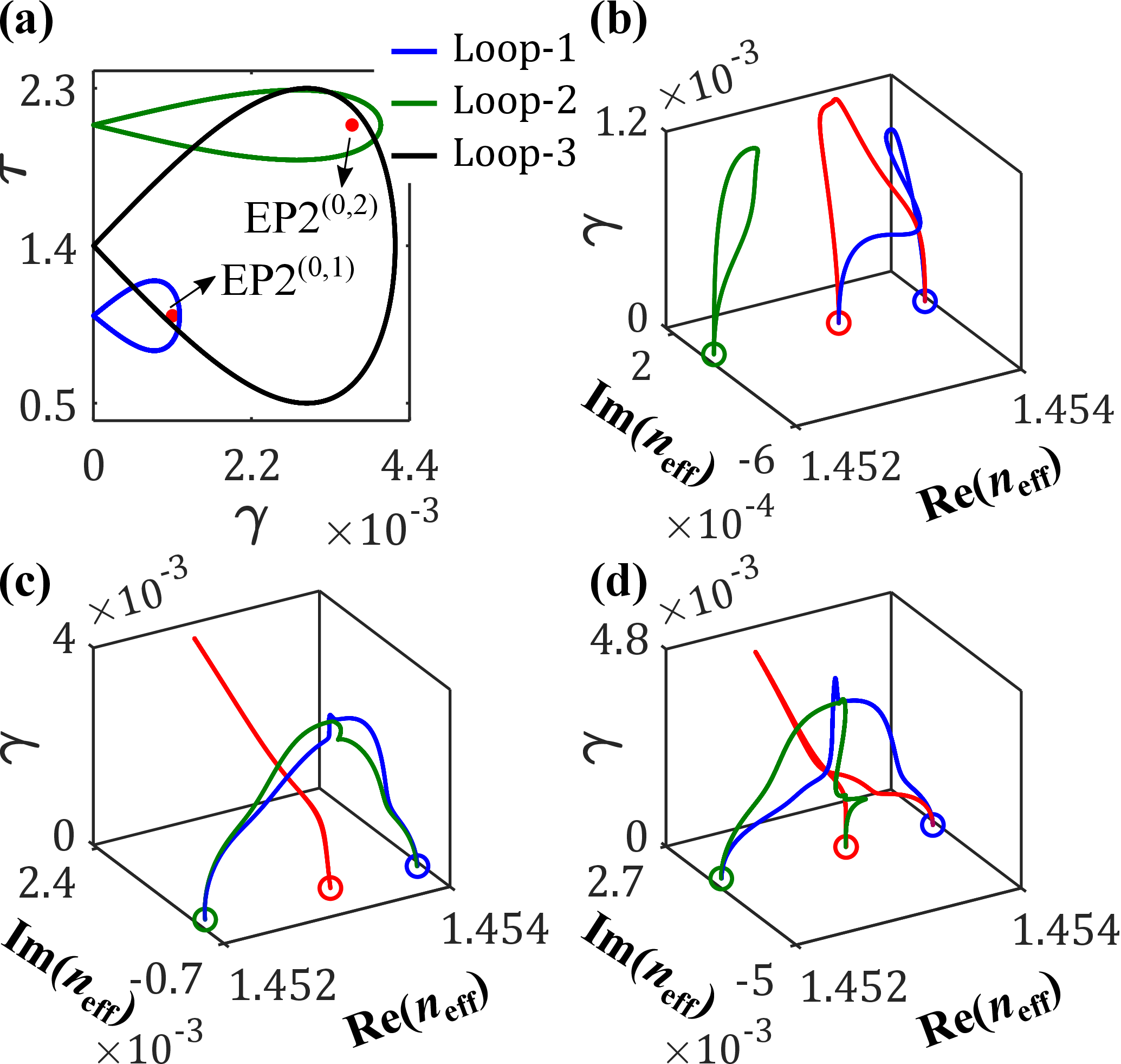}
	\caption{\textbf{(a)} Three chosen parametric loops in the $(\gamma,\tau)$-plane to investigate the topological properties of second- and third-order branch points. Two red dots indicate the coordinates of two interconnected EP2s. Trajectories of  $n_{\text{eff}}$-values associated with $\Psi_0$, $\Psi_1$, and $\Psi_2$ (represented by blue, red, and green lines, respectively), while considering a clockwise (CW) quasistatic variations of $\gamma$ and $\tau$ \textbf{(b)} along Loop-1, exhibiting the second order branch point behavior of EP2$^{(0,1)}$ with permutations $n_{\text{eff}}(\Psi_0)\rightarrow n_{\text{eff}}(\Psi_1)\rightarrow n_{\text{eff}}(\Psi_0)$ and $n_{\text{eff}}(\Psi_2)\rightarrow n_{\text{eff}}(\Psi_2)$; \textbf{(c)} along Loop-2, exhibiting the second order branch point behavior of EP2$^{(0,2)}$ with permutations $n_{\text{eff}}(\Psi_0)\rightarrow n_{\text{eff}}(\Psi_2)\rightarrow n_{\text{eff}}(\Psi_0)$ and $n_{\text{eff}}(\Psi_1)\rightarrow n_{\text{eff}}(\Psi_1)$; \textbf{(c)} along Loop-3, exhibiting the third order branch point behavior of an analogous EP3 formed by the interconnected EP2$^{(0,1)}$ and EP2$^{(0,2)}$ with permutations $n_{\text{eff}}(\Psi_0)\rightarrow n_{\text{eff}}(\Psi_2)\rightarrow n_{\text{eff}}(\Psi_1)\rightarrow n_{\text{eff}}(\Psi_0)$. In (b)-(d), the circular markers of respective colors represent the starting positions of $n_{\text{eff}}$-values. Here, each point of progression along the trajectory of the $n_{\text{eff}}$-value for each mode is precisely synchronized with a corresponding point of progression on a distinct loop in the $(\gamma,\tau)$-plane.}
	\label{fig_3}
\end{figure}

Figure \ref{fig_3}(a) displays the coordinates of EP2$^{(0,1)}$ and EP2$^{(0,2)}$, along with three distinct encirclement schemes based on Eq. \eqref{eq_enc} within the $(\gamma,\tau)$-plane. The topological properties of second-order branch points can be examined by encircling each EP2 individually. However, the emergence of an EP3, with its characteristic third-order branch point behavior, becomes evident when a parametric loop encloses both EP2s simultaneously. Accordingly, we define three specific encirclement schemes as follows: Loop-1 (blue loop), with parameters $\gamma_0=1.2\times10^{-3}$, $\tau_0=1$, and $r=0.4$, encloses only EP2$^{(0,1)}$; Loop-2 (green loop), defined by $\gamma_0=4.2\times10^{-3}$, $\tau_0=2.008$, and $r=0.2$, encircles only EP2$^{(0,2)}$; and Loop-3 (black loop), characterized by $\gamma_0=4.2\times10^{-3}$, $\tau_0=1.4$, and $r=0.9$, encloses both the EP2s together.

The consequences of the quasistatic encirclement scheme along the selected loops and the underlying topological behaviors are analyzed by tracing the corresponding trajectories of the $n_{\text{eff}}$-values associated with $\Psi_j\,(j=0,1,2)$, as depicted in Figs. \ref{fig_3}(b)--\ref{fig_3}(d). The trajectories are represented by dotted blue, red, and green curves, respectively, where three circular markers of the same colors indicates their starting points [i.e., when $\theta=0$ in Eq. \eqref{eq_enc}]. Each point along these trajectories in the complex $n_{\text{eff}}$-plane (where $\gamma$ is plotted along an additional axis) corresponds to a unique point on a particular loop in the $(\gamma,\tau)$-plane.

While considering a complete $2\pi$ rotation in the clockwise (CW) direction along Loop-1 [allowing a quasistatic gain-loss variation around only EP2$^{(0,1)}$, while keeping EP2$^{(0,2)}$ outside], we can observe an adiabatic swapping between the $n_{\text{eff}}$-values associated with the modes connected through EP2$^{(0,1)}$, i.e., $\Psi_0$ and $\Psi_1$, [like, $n_{\text{eff}}(\Psi_0)\rightarrow n_{\text{eff}}(\Psi_1)\rightarrow n_{\text{eff}}(\Psi_0)$] in Fig. \ref{fig_3}(b). Meanwhile, the $n_{\text{eff}}$-value associated with $\Psi_2$ remains unchanged [$n_{\text{eff}}(\Psi_2)\rightarrow n_{\text{eff}}(\Psi_2)$; as also evident in \ref{fig_3}(b)], indicating it is unaffected by the structured gain-loss modulation along Loop-1. In a similar fashion, a complete $2\pi$ CW rotation along Loop-2 [allowing a quasistatic gain-loss variation around only EP2$^{(0,2)}$, while keeping EP2$^{(0,1)}$ outside] results in an adiabatic swapping between the $n_{\text{eff}}$-values of $\Psi_0$ and $\Psi_2$, leaving the $n_{\text{eff}}$-value of $\Psi_1$ unchanged [like, $n_{\text{eff}}(\Psi_0)\rightarrow n_{\text{eff}}(\Psi_2)\rightarrow n_{\text{eff}}(\Psi_0)$; $n_{\text{eff}}(\Psi_1)\rightarrow n_{\text{eff}}(\Psi_1)$], as shown in Fig. \ref{fig_3}(c). Such intriguing interactions among the $n_{\text{eff}}$-values of three coupled modes, as observed in Figs. \ref{fig_3}(b) and \ref{fig_3}(c), which display distinct adiabatic permutations associated with the pairs $\{\Psi_0,\Psi_1\}$ and $\{\Psi_0,\Psi_2\}$, reveal the individual second-order branch-point topology of EP2$^{(0,1)}$ and EP2$^{(0,2)}$. Notably, with an anticlockwise (ACW) parametric variation along both Loop-1 and Loop-2, similar $n_{\text{eff}}$ trajectories are observed, differing only in that the two swapping modes exchange their paths.

Now, we consider a quasistatic variation of $\gamma$ and $\tau$ along Loop-3, which simultaneously encloses both EP2$^{(0,1)}$ and EP2$^{(0,2)}$. Such a patterned perturbation results in a successive and adiabatic permutation among the $n_{\text{eff}}$-values of all the coupled modes. Figure \ref{fig_3}(d) shows the results for a CW encirclement process along Loop-3, where we can observe the permutation pattern $n_{\text{eff}}(\Psi_0)\rightarrow n_{\text{eff}}(\Psi_2)\rightarrow n_{\text{eff}}(\Psi_1)\rightarrow n_{\text{eff}}(\Psi_0)$. However, we can observe a different pattern like $n_{\text{eff}}(\Psi_0)\rightarrow n_{\text{eff}}(\Psi_1)\rightarrow n_{\text{eff}}(\Psi_2)\rightarrow n_{\text{eff}}(\Psi_0)$ upon considering the ACW encirclement process along Loop-3. Such characteristic features of $n_{\text{eff}}$ trajectories vividly illustrates the topology of a third-order branch point and demonstrates the emergence of an EP3 in the presence of interconnected EP2s within the same 2D $(\gamma,\tau)$-plane.

\subsection{Effect of dynamical parametric variation: Consideration of nonadiabatic terms}

Now, we study the propagation of the quasiguided modes $\Psi_j\,(j=0,1,2)$, while considering dynamic variation of the control parameters $\gamma$ and $\tau$. To achieve this, we tailor the spatial gain-loss distribution [i.e., essentially the Im($n$) profile] defined by Eq. \eqref{eq_enc} along the fiber length (i.e., along $z$-axis). Accordingly, we substitute $\theta = 2\pi z / L$ in Eq. \eqref{eq_enc} to map $\theta = \{0, 2\pi\}$ to $z = \{0, L\}$. This substitution leads to the dynamic parameter distribution:
\begin{align}
	&\gamma(x,y,z)=\gamma_{0}\sin\left(\frac{\pi z}{L}\right),\notag\\
	&\tau (x,y,z)=\tau_{0}+r\sin\left(\frac{2\pi z}{L}\right).
	\label{eq_denc}
\end{align}
Here, $\gamma$ and $\tau$ vary solely along the $z$-axis; they remain fixed across any cross-section in the $xy$-plane of the fiber. Equation \eqref{eq_denc} implies that one complete pass of light through the fiber ($z:0\rightarrow L$) corresponds to a full cycle in the parametric loop ($\theta:0\rightarrow 2\pi$). Here CW parametric variation ($\theta:0\rightarrow 2\pi$) can be realized by considering the light propagation in the forward direction with input at $z=0$ and output at $z=L$. Conversely, ACW parametric variation ($\theta:2\pi\rightarrow 0$) can be achieved by reversing the propagation direction (i.e., along the backward direction) with input at $z=L$ and output at $z=0$. Notably, Eq. \eqref{eq_denc} allows for $\gamma = 0$ at both $z = 0$ and $z = L$, which facilitates the excitation and retrieval of passive modes in both propagation directions. This would not be as straightforward for other parametric loop shapes. 

To analyze EP-induced light dynamics with adiabatically expected mode conversions, we must account for implications of the adiabatic theorem, which introduces certain nonadiabatic correction terms. These terms play a crucial role when considering dynamic parametric variations. Based on the time-dependent Schr\"odinger equation associated with a Hamiltonian (in the quantum formalism) \cite{Gilary13na}, we can express an optical counterpart for the key proportional factors associated with nonadiabatic corrections in the proposed fiber structure as 
\begin{subequations}
\begin{align}
&\mathcal{N}_{m\rightarrow n}\propto-\exp\left[\displaystyle\int_0^L\Delta\Gamma_{m,n}(\gamma,\tau)dz\right]\label{NA1}\\
&\mathcal{N}_{n\rightarrow m}\propto+\exp\left[\displaystyle\int_0^L\Delta\Gamma_{m,n}(\gamma,\tau)dz\right]\label{NA2}
\end{align}
\label{NA}%
\end{subequations}
In Eq. \eqref{NA}, the indices $\{m,n\}$ signify the all possible transitions among $\Psi_j\,(j=0,1,2)$ [i.e., $\{m,n\}\in j;\,m\ne n$], where $m\rightarrow n$ and $n\rightarrow m$ in Eqs.\eqref{NA1} and \eqref{NA2} correspond to the transitions $\Psi_m\rightarrow\Psi_n$ and $\Psi_n\rightarrow\Psi_m$, respectively (as expected adiabatically). 

In Eq. \eqref{NA}, the factor $\Delta\Gamma_{m,n}$, known as the relative gain, plays the key role in adiabatic breakdown in the EP-based light dynamics. The relative gain $\Delta\Gamma_{m,n}$ is defined as the relative difference between the average loss ($\Gamma_{\text{av}}$) experienced by each mode. To estimate $\Gamma_{\text{av}}$ for a mode, we can consider the corresponding adiabatic trajectory of Im($n_{\text{eff}}$) during a given encirclement scheme, where $\Gamma_{\text{av}}=(2\pi)^{-1}\oint\text{Im}(n_{\text{eff}})d\theta$ (it approximates the mode's accumulated loss over one cycle). Depending on the adiabatic evolution of $n_{\text{eff}}$-values for a given encirclement scheme, two distinct cases arise: either $\Delta\Gamma_{m,n}>0$ or $\Delta\Gamma_{m,n}<0$. These relations determine whether nonadiabatic corrections terms ($\mathcal{N}$) either dominate or support the anticipated adiabatic dynamics, influenced by the corresponding amplifying or decaying exponential terms in Eq. \eqref{NA}. For instance, when $\Delta\Gamma_{m,n}>0$, $\mathcal{N}_{m\rightarrow n}$ converges due to a decaying exponential term, while $\mathcal{N}_{n\rightarrow m}$ diverges due to an amplifying exponential term. This results in the validation of the adiabaticity for the transition $\Psi_m\rightarrow\Psi_n$, whereas violation of the adiabaticity for the transition $\Psi_n\rightarrow\Psi_m$. Consequently, $\Psi_m$ evolves adiabatically, whereas $\Psi_n$ undergoes a nonadiabatic transition. On the other hand, for $\Delta\Gamma_{m,n}<0$, we can similarly estimate the nonadiabatic correction terms from Eq. \eqref{NA}, where we obtain an exactly opposite scenario: $\Psi_n$ evolves adiabatically, while $\Psi_m$ follows a nonadiabatic transition.

In the following sections, we validate our prior analysis of EP-induced light dynamics by examining mode propagation characteristics obtained using the Beam Propagation Method (BPM) in RSoft\textsuperscript{\textregistered} simulation software. Specifically, we explore the distinctive features of both chiral and non-chiral light dynamics for various dynamical encirclement schemes.  

\subsection{Characteristics of chiral light dynamics}

\begin{figure}[htpb]
	\centering
	\includegraphics[width=\linewidth]{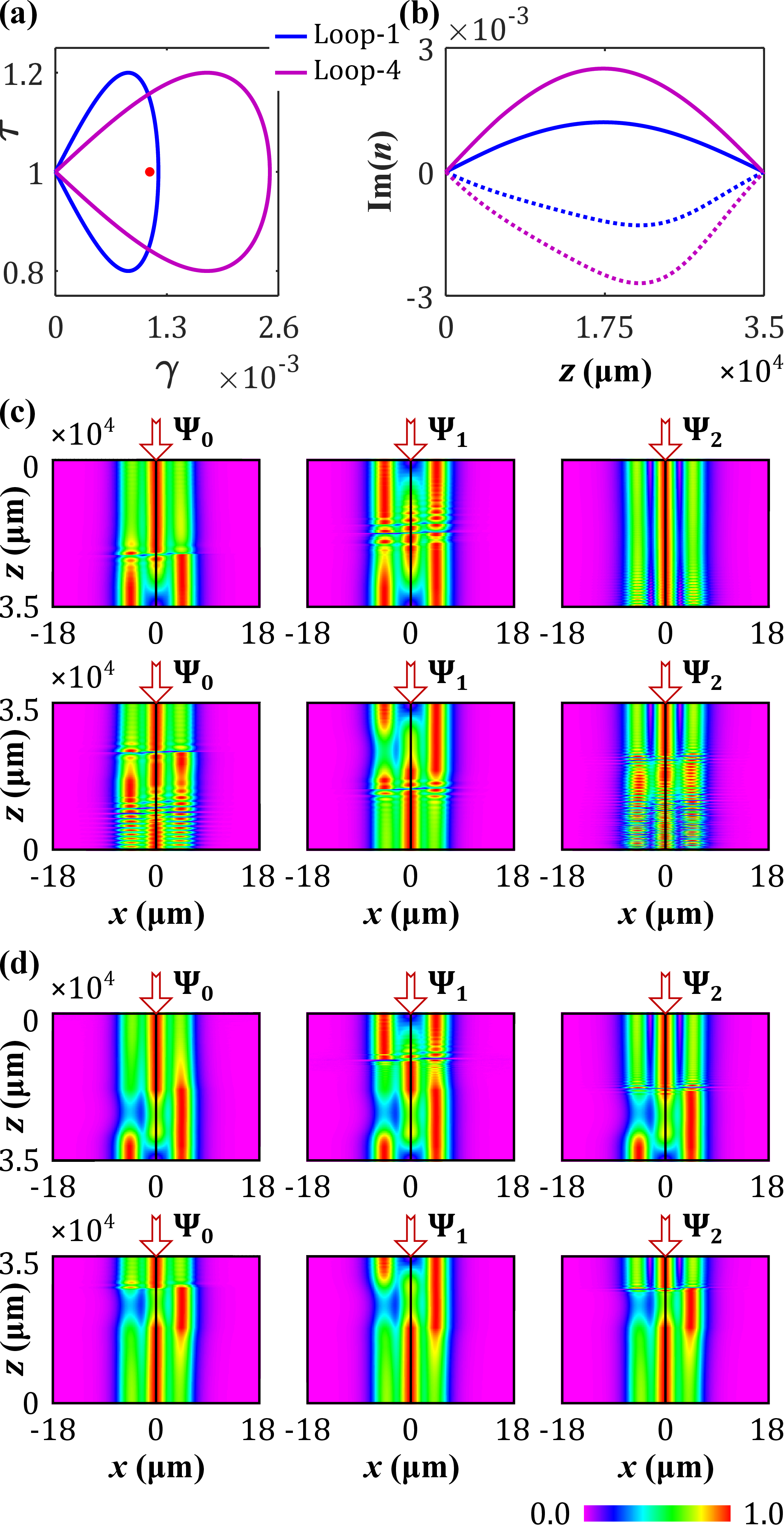}
	\caption{\textbf{(a)} Loop-1 (blue loop) and Loop-4 (violet loop), enclosing only EP2$^{(0,1)}$ (red dot), along with \textbf{(b)} corresponding mapped length-dependent distributions of gain (dotted lines of respective colors) and loss (solid lines of respective colors) in the leftmost and rightmost cores. Beam propagation simulation results while dynamically varying the control parameters \textbf{(c)} along Loop-1 in (upper panel) the CW direction ($z:0\rightarrow L$), resulting in asymmetric conversions $\{\Psi_0,\Psi_1\}\rightarrow\Psi_1$ and $\Psi_2\rightarrow\Psi_2$, and in (lower panel) the ACW direction ($z:L\rightarrow 0$), resulting in asymmetric conversions $\{\Psi_0,\Psi_1\}\rightarrow\Psi_0$ and $\Psi_2\rightarrow\Psi_2$; \textbf{(d)} along Loop-4 in (upper panel) the CW direction ($z:0\rightarrow L$), resulting in asymmetric conversions $\{\Psi_0,\Psi_1,\Psi_2\}\rightarrow\Psi_1$, and in (lower panel) the ACW direction ($z:L\rightarrow 0$), resulting in asymmetric conversions $\{\Psi_0,\Psi_1,\Psi_2\}\rightarrow\Psi_0$. We re-normalize the intensities at each $z$ during propagation to accurately illustrate the evolution of the modes.}
	\label{fig_4}
\end{figure}

Figure \ref{fig_4} presents the beam propagation simulation results under a dynamical encirclement scheme only around EP2$^{(0,1)}$, where EP2$^{(0,2)}$ remains away from the encirclement regime. Initially, we examine the beam dynamics for Loop-1 (blue loop), as depicted in Fig. \ref{fig_4}(a). In Fig. \ref{fig_4}(b), the solid and dotted blue lines represent the length-dependent variations in loss and gain [after mapping Loop-1 via Eq. \eqref{eq_denc}] in the rightmost and leftmost cores, respectively. For this setup, when a CW dynamical encirclement scheme is considered by exciting the modes from $z=0$, both modes $\Psi_0$ and $\Psi_1$ [connected through EP2$^{(0,1)}$] are converted into $\Psi_1$ at $z=L$. Here, the nonadiabatic correction terms appear with $\Delta\Gamma_{0,1}>0$, indicating that $\Psi_0$ undergoes an adiabatic transition, while $\Psi_1$ experiences a nonadiabatic transition. Meanwhile, $\Psi_2$ remains unaffected by the dynamical parametric variation along Loop-1, retaining in $\Psi_2$ at $z=L$. Such asymmetric conversion of modes [$\{\Psi_0,\Psi_1\}\rightarrow\Psi_1$ and $\Psi_2\rightarrow\Psi_2$] are shown in the upper panel of Fig. \ref{fig_4}(c). Conversely, under an ACW dynamical encirclement scheme by exciting the modes from $z=L$, $\Psi_0$ and $\Psi_1$ are converted convert into $\Psi_0$ at $z=0$. Here, the nonadiabatic correction terms appear with $\Delta\Gamma_{0,1}<0$. Hence, $\Psi_0$ and $\Psi_1$ undergo nonadiabatic and adiabatic transitions, respectively, while $\Psi_2$ remains unchanged, leading to the asymmetric mode conversions $\{\Psi_0,\Psi_1\}\rightarrow\Psi_0$ and $\Psi_2\rightarrow\Psi_2$, as shown in the lower panel of Fig. \ref{fig_4}(c). 

It is noteworthy that the gain-loss accumulated over Loop-1 is insufficient to impact the nonadiabatic correction terms associated with $\Psi_2$, leaving it unaffected in both propagation directions, as shown in Fig. \ref{fig_4}(c). However, $\Psi_2$ would no longer remain unaffected if we consider a comparatively larger loop that still encircles only EP2$^{(0,1)}$. We examine such a scenario by considering a new loop (say, Loop-4, defined by parameters $\gamma_0 = 2.5 \times 10^{-3}$, $\tau_0 = 1$, and $r=0.4$), as illustrated by the violet loop in Fig. \ref{fig_4}(a). The solid and dotted violet lines in Fig. \ref{fig_4}(b) show the corresponding mapped loss and gain distribution, respectively, in the fiber structure. By comparing the gain-loss profiles induced by Loop-1 and Loop-4, we observe that the accumulated gain-loss in Loop-4 is considerably higher. This increased accumulation gain-loss brings $\Psi_2$ into consideration for interaction induced by EP2$^{(0,1)}$, due to the involvement of associated nonadiabatic correction factors $\Delta\Gamma_{0,2}$ and $\Delta\Gamma_{1,2}$. The corresponding beam propagation results are shown in Fig. \ref{fig_4}(d). In the upper panel, the effective nonadiabatic corrections based on $\Delta\Gamma_{1,2} < 0$ and $\Delta\Gamma_{0,1} > 0$ during a CW parametric variation along Loop-4 lead all three modes $\Psi_j , (j = 0, 1, 2)$ to convert into $\Psi_1$, resulting in $\{\Psi_0, \Psi_1, \Psi_2\} \rightarrow \Psi_1$. In contrast, the ACW parametric variation along Loop-4 introduces effective nonadiabatic corrections based on $\Delta\Gamma_{0,2} < 0$ and $\Delta\Gamma_{0,1} < 0$, resulting the asymmetric conversion $\{\Psi_0, \Psi_1, \Psi_2\} \rightarrow \Psi_0$, as illustrated in the lower panel of Fig. \ref{fig_4}(d).

Therefore, while considering the dynamical parameter space defined by Loop-1, light dynamics is partially driven by the device's chirality. Specifically, the transitions of $\Psi_0$ and $\Psi_1$ yield direction-dependent outputs, regardless of the input, while $\Psi_3$ remains unaffected in both propagation directions [as illustrated in Fig. \ref{fig_4}(c)]. In contrast, with the parametric variation along Loop-4, fully chirality-driven asymmetric light dynamics is observed, where regardless of the initial mode, light is converted into different dominant modes depending on the propagation direction. Thus, for multimode structures, the overall chiral dynamics depends not only on whether the relevant EP2 is enclosed within the parametric loop but also on the total accumulated gain-loss influencing interactions across all supported modes.

\subsection{Characteristics of nonchiral light dynamics}

\begin{figure}[t!]
	\centering
	\includegraphics[width=\linewidth]{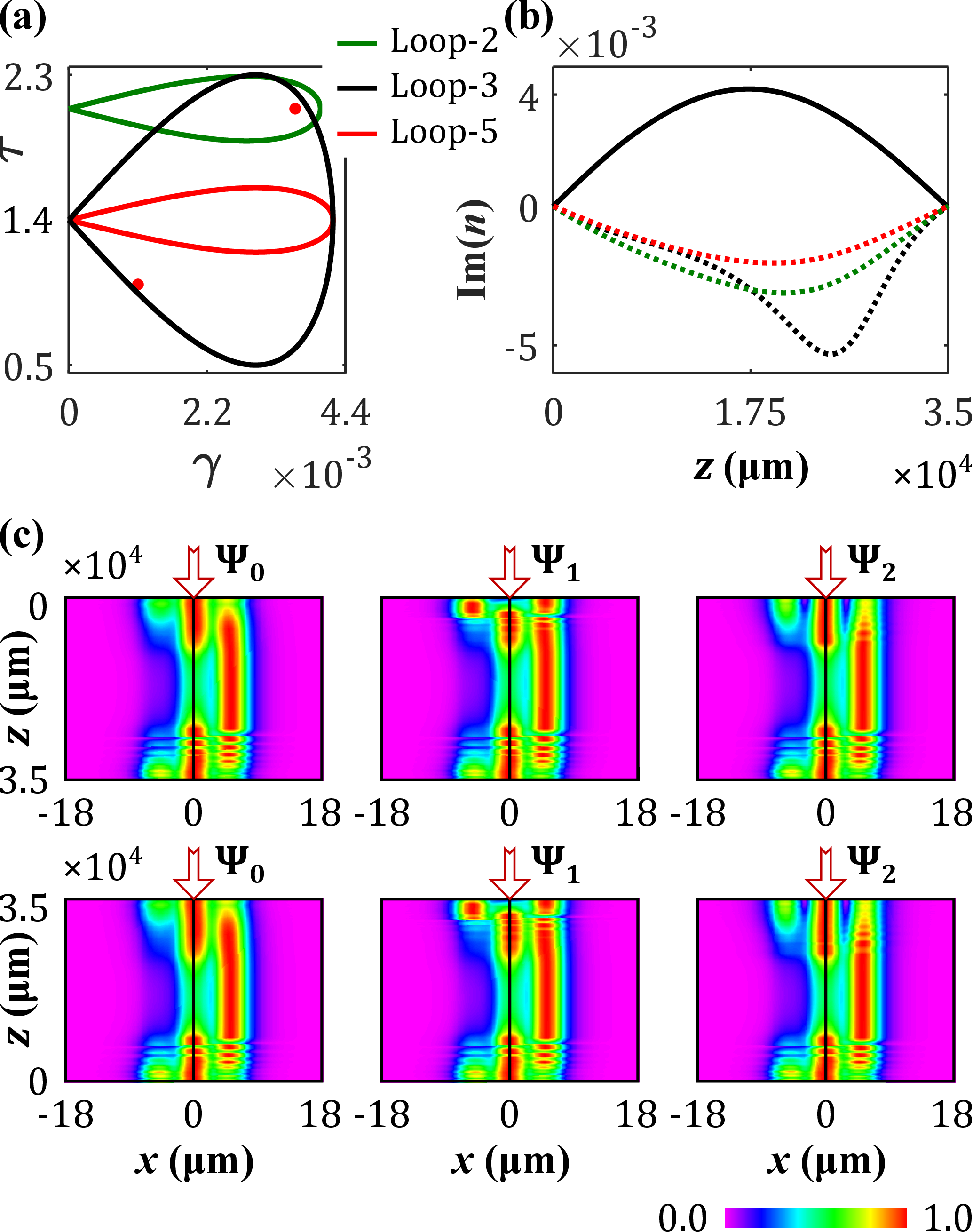}
	\caption{\textbf{(a)} Loop-2 (green loop), Loop-3 (black loop), and Loop-5 (red loop) in the proximity of EP2$^{(0,1)}$ and EP2$^{(0,2)}$ (two red dots). \textbf{(b)} Corresponding mapped length-dependent distributions of gain (dotted lines of respective colors) and loss (solid black line, as the loss distribution is same for all the three mentioned loops) in the leftmost and rightmost cores. \textbf{(c)} Beam propagation simulation results while dynamically varying the control parameters along each of the chosen loops (individually) in (upper panel) the CW direction ($z:0\rightarrow L$) and (lower panel) the ACW direction ($z:L\rightarrow 0$), resulting in nonchiral conversions $\{\Psi_0,\Psi_1,\Psi_2\}\rightarrow\Psi_0$ (for both the directions). Note that, we observe similar beam propagation dynamics for all three mentioned loops; therefore, results are presented for only one case. We re-normalize the intensities at each $z$ during propagation to accurately illustrate the evolution of the modes.}
	\label{fig_5}
\end{figure}

In Fig. \ref{fig_5}, we analyze three another scenarios based on three distinct parametric loops, where for each of the cases a breakdown of the chiral behavior can be revealed. Two of these scenarios involve previously considered configurations: Loop-2, which encloses only EP2$^{(0,2)}$, and Loop-3, which encircles an analogous EP3 along with its connected EP2$^{(0,1)}$ and EP2$^{(0,2)}$. The third case introduces a new parametric loop, say Loop-5 (defined by parameters $\gamma_0 = 4.2 \times 10^{-3}$, $\tau_0 = 1.4$, and $r = 0.2$), which passes close to both EP2s without encircling them. Figure \ref{fig_5}(a) illustrates these three loops, where their corresponding mapped parameter spaces, characterizing the length-dependent loss and gain distributions in the rightmost and leftmost cores, are presented in Fig. \ref{fig_5}(b). Notably, under the specified characteristic parameters ${\gamma_0, \tau_0, r}$ for the proposed setup [primarily defined by Eqs. \eqref{eq_sys} and \eqref{eq_denc}], the loss distribution in the rightmost core remains identical across all three loops, as shown by the solid black line. The dotted lines of corresponding colors indicate the gain distributions in the leftmost core for each of the loops. As a result, the gain-loss contrasts vary among these three loops.

We begin by considering dynamical encirclement along Loop-2, where the nonadiabatic correction factors exhibit unusual behavior. Although Loop-2 encloses only EP2$^{(0,2)}$ (associated with modes $\Psi_0$ and $\Psi_2$) the accumulated gain-loss over Loop-2 also takes $\Psi_1$ into account for interactions, necessitating the inclusion of nonadiabatic correction factors associated with all three modes to accurately capture the actual dynamics of light. Here, for a CW dynamical encirclement scheme (with input at $z=0$ and output at $z=L$), we obtain $\Delta\Gamma_{0,2} > 0$, $\Delta\Gamma_{0,1} < 0$ and $\Delta\Gamma_{1,2} < 0$. Now, $\Delta\Gamma_{0,1} < 0$ essentially allows the mode conversions $\{\Psi_0,\Psi_1\}\rightarrow\Psi_0$. Additionally, owing to $\Delta\Gamma_{0,2} > 0$ both $\Psi_0$ and $\Psi_2$ are expected to convert into $\Psi_2$, however, due to an additional combined effect of $\Delta\Gamma_{1,2} < 0$ and $\Delta\Gamma_{0,1} < 0$ both $\Psi_0$ and $\Psi_2$ ultimately end up at $\Psi_0$. Thus, we observe the overall mode conversions $\{\Psi_0,\Psi_1,\Psi_2\}\rightarrow\Psi_0$ during propagation from $z=0$ to $z=L$, as exactly demonstrated by the beam propagation simulation results in the upper panel of Fig. \ref{fig_5}(c). For an ACW dynamical encirclement scheme (with input at $z=L$ and output at $z=0$), the nonadiabatic correction factors based on $\Delta\Gamma_{0,2} < 0$ and $\Delta\Gamma_{0,1} < 0$ enable similar mode conversions. The resulting transitions mirror the CW case, leading to $\{\Psi_0, \Psi_1, \Psi_2\}\rightarrow\Psi_0$, as shown in the lower panel of Fig. \ref{fig_5}(c). Therefore, the overall light dynamics along Loop-2 is nonchiral, where, irrespective of both inputs and direction of propagation, light is converted into a particular dominant mode. Comparing with adiabatic expectations [as shown in Fig. \ref{fig_3}(c)], we observe that only $\Psi_2$ follows an adiabatic transition, whereas $\Psi_0$ and $\Psi_1$ undergo nonadiabatic transitions in both directions.

Now, we study the effect of dynamical encirclement around an analogous EP3 formed with two interconnected EP2s. Loop-3 allows us to consider such a situation by encircling both EP2$^{(0,1)}$ and EP2$^{(0,2)}$ simultaneously. Here, for a CW dynamical encirclement scheme (with input at $z=0$ and output at $z=L$), we obtain $\Delta\Gamma_{0,1} < 0$, $\Delta\Gamma_{0,2} < 0$, and $\Delta\Gamma_{1,2} < 0$. Now, the first two relations are sufficient to understand the overall light dynamics. These two relations enable the asymmetric conversion of $\Psi_1$ and $\Psi_2$ into $\Psi_0$. Additionally, both relations validate the retention of $\Psi_0$ into $\Psi_0$. Interestingly, the combination of the first and third relations also supports the same output. While the third relation suggests that both $\Psi_1$ and $\Psi_2$ are expected to convert into $\Psi_1$, they ultimately end up at $\Psi_0$ due to the influence of the first relation. Hence, during the propagation from $z=0$ to $z=L$, light is finally converted into the dominant $\Psi_0$, with the conversions $\{\Psi_0, \Psi_1, \Psi_2\} \rightarrow \Psi_0$, as similar to the beam propagation results shown in the upper panel of Fig. \ref{fig_5}(c). On the other hand, for an ACW dynamical encirclement scheme (with input at $z=L$ and output at $z=0$), the nonadiabatic correction factors read the relations $\Delta\Gamma_{0,1} < 0$, $\Delta\Gamma_{0,2} < 0$, and $\Delta\Gamma_{1,2} > 0$. Here, the first two relations are similar to the CW case, whereas the third one is opposite to the CW case. Consequently, based on the associated nonadiabatic correction factors, all three modes are converted into $\Psi_0$ while propagating from $z=L$ to $z=0$, as shown in the beam propagation results in the lower panel of Fig. \ref{fig_5}(c). Therefore, the overall light dynamics along Loop-3, i.e., around an EP3, is also nonchiral. Comparing with adiabatic expectations [as shown in Fig. \ref{fig_3}(d)], we observe that only $\Psi_1$ follows an adiabatic transition in the forward direction, whereas $\Psi_2$ follows an adiabatic transition in the backward direction. The rest of the transitions are nonadiabatic.

Furthermore, we investigate the light dynamics under the dynamical variations of $\gamma$ and $\tau$ along Loop-5. This loop traverses the interaction regimes of both EP2$^{(0,1)}$ and EP2$^{(0,2)}$, passing close to these points without enclosing either. This configuration is selected to test the recent claim of asymmetric light dynamics occurring without enclosing EP2s \cite{Hassan2017,Nasari2022}, particularly to verify if this behavior extends to higher-order EPs. Based on the characteristics of the associated $n_{\text{eff}}$-values, it is expected that all modes will remain in their original states without any asymmetric conversion. However, it is crucial to examine whether associated nonadiabatic factors could introduce unexpected behavior in the light dynamics. For Loop-5, we observe relationships between the relative gain factors similar to those found for Loop-3. Specifically, for a CW variation, we have $\Delta\Gamma_{0,1} < 0$, $\Delta\Gamma_{0,2} < 0$, and $\Delta\Gamma_{1,2} < 0$, while for an ACW variation, the relationships are $\Delta\Gamma_{0,1} < 0$, $\Delta\Gamma_{0,2} < 0$, and $\Delta\Gamma_{1,2} > 0$. The influence of associated nonadiabatic correction factors enables the nonchiral conversions $\{\Psi_0, \Psi_1, \Psi_2\} \rightarrow \Psi_0$, irrespective of the input modes and propagation directions, consistent with the beam propagation results illustrated in Fig. \ref{fig_5}(c). Therefore, the light dynamics retains its nonchiral behavior for Loop-5, similar to the case of Loop-3; however, in this scenario, all transitions are classified as nonadiabatic.

The key question now is how long the characteristics of asymmetric mode conversion induced by a dynamically encircled EP3 remain evident when any of the connected EP2s are note enclosed. To investigate this, we gradually reduce the size of Loop-3 by decreasing the parameter $r$ (starting from $r=0.9$) in Eq. \eqref{eq_enc}, while keeping the other parameters, $\gamma_0$ and $\tau_0$, fixed. We then examine the behavior of relative gain factors and associated beam propagation results to determine if they remain consistent with the original Loop-3 case. Loop-5 represents the optimized parameter space with $r=0.2$ until which the nonadiabatic correction factors behave similarly to those in Loop-3. Below this threshold (for the proposed fiber structure), asymmetric conversions are no longer observed, and all modes remain in their original states while transitioning. It is important to note that loops with similar characteristics can also be defined by varying $\gamma_0$ or $\tau_0$ (instead of only varying $r$). In general, our observations suggest that as long as the nonadiabatic correction factors retain properties analogous to those observed during dynamical encirclement enclosing connected EP2s (i.e., an analogous EP3), the phenomenon of asymmetric mode conversion persists, even without enclosing an EP3 or the two connected EP2s.

Although similar nonchiral light dynamics are observed for Loop-2, Loop-3, and Loop-5, Loop-5 accumulates a significantly lower amount of gain with reduced gain-loss contrast. This observation suggests that it is possible to design a parametric loop as small as feasible, minimizing both the accumulated gain and the gain-loss contrast, while still achieving equivalent light dynamics. Such a configuration offers enhanced fabrication feasibility, making it more practical for implementation.

\section{Conclusions}

This study explores the advanced light manipulation capabilities enabled by higher-order EPs in a non-Hermitian photonic system with circular geometry. Focusing on a custom-engineered triple-core optical fiber, we investigate the topological and dynamic properties of an EP3 formed by two interconnected EP2s. We consider different parametric loops through tailored gain-loss modulation in a 2D parameter space to reveal various chiral and nonchiral aspects of light dynamics assisted with asymmetric mode conversions. We observe that the chiral or nonchiral aspects depend not only on whether the relevant EP2 is enclosed within the parametric loop but also on the total accumulated gain-loss influencing interactions across all supported modes and, thereby, the influence of associated nonadiabatic corrections. Notably, this research demonstrates that the effects induced by higher-order EPs extend beyond direct encirclement, persisting in regions near but not enclosing the connected EP2s, thereby facilitating efficient and compact designs for practical applications. Overall, the results showcase a simplified approach to leverage non-Hermiticity for hosting the topological and dynamic properties of higher-order EPs without requiring complex parameterization, demonstrating precise control over light dynamics. Our findings pave the way for fiber-based higher-order mode converters with improved mode selectivity and multi-modal functionality, addressing critical needs in all-fiber communication networks and signal processing. Furthermore, the proposed concept offers significant potential for experimental realization, with promising applications in developing all-fiber components such as isolators, circulators, and ultra-sensitive optical fiber sensors.

\begin{acknowledgments}
A.L. and A.M. acknowledge the financial support from the Maestro Grant (No. DEC-2019/34/A/ST2/00081) of the Polish National Science Center (NCN).
\end{acknowledgments}

\bibliography{References}

\end{document}